\documentclass{cup-hpl}

\usepackage[utf8]{inputenc}

\usepackage{xcolor}

\usepackage{graphicx}
\usepackage{hyperref}
\usepackage{siunitx}
\DeclareMathSymbol{\Omega}{\mathalpha}{operators}{10}
\usepackage[title]{appendix}

\begin{document}

\shorttitle{HRR source of \si{\nano\second} duration \si{\kilo\ampere} pulses}                                   
\shortauthor{M.~Ehret, J.~Cikhardt, P.~Bradford, et al.}

\title{{High-repetition-rate source of nanosecond duration\\ \si{\mathbf{\kilo\ampere}}-current pulses driven by relativistic laser pulses}}



\author[1]{Michael~Ehret\corresp{C Adaja 8, ES-37185, or \email{mehret@clpu.es}}}

\author[2]{Jakub~Cikhardt}

\author[3]{Philip~Bradford}

\author[1]{Iuliana-Mariana~Vladisavlevici}

\author[4]{Tomas~Burian}

\author[1]{Diego~de~Luis}

\author[1]{Jose~Luis~Henares}

\author[1]{Rub\'{e}n~Hern\'{a}ndez~Mart\'{i}n}

\author[1]{Jon~Imanol~Api\~{n}aniz}

\author[1]{Roberto~Lera}

\author[1]{Jos\'{e}~Antonio~P\'{e}rez-Hern\'{a}ndez}

\author[3]{Jo\~{a}o~Jorge~Santos}

\author[1]{Giancarlo~Gatti}

\address[1]{Centro de L\'{a}seres Pulsados (CLPU), Villamayor, Spain}
\address[2]{Czech Technical University in Prague, Faculty of Electrical Engineering, Prague, Czech Republic}
\address[3]{Univ. Bordeaux-CNRS-CEA, Centre Lasers Intenses et Applications (CELIA), UMR 5107, Talence, France}
\address[4]{Department of Radiation and Chemical Physics, FZU-Institute of Physics of the Czech Academy of Sciences, Prague, Czech Republic}

\date{\today}

\begin{abstract}
We report the {first} high-repetition rate generation and {simultaneous} characterization of nanosecond-scale return currents of {\si{\kilo\ampere}-magnitude} issued by the {polarization} of a target irradiated with a PW-class high repetition rate Ti:Sa laser system {at relativistic intensities. We present experimental results obtained with the VEGA-3 laser at intensities from \SIrange{5e18}{1.3e20}{\watt\per\square\centi\metre}}. {A} {non-invasive} inductive return current monitor is adopted to measure the derivative of return-currents on the order of \si{\kilo\ampere\per\nano\second} and analysis methodology is developed to derive return-currents. We compare the current for copper, aluminium and Kapton targets at different laser energies. The data shows the stable production of current peaks {and clear prospects for the tailoring of the pulse shape,} promising for future applications {in high energy density science}, {e.g. electromagnetic interference stress tests, high-voltage pulse response measurements, and charged particle beam lensing}. {We compare the} target discharge of the order of hundreds of \si{\nano\coulomb} with theoretical predictions {and a good agreement is found.}
\end{abstract}


\maketitle

\section{Introduction}

The continuous technical and scientific improvement of lasers \cite{MAIMAN1966,DiDomenico1966} {has} led to stable short-pulse PW-class high repetition rate Ti:Sa systems \cite{Maine1988,Aoyama2003}. If {these lasers are} tightly focused onto matter, the relativistic interaction yields forward-acceleration of electrons \cite{TajimaMalka2020} that in turn can trigger pulsed bright ion beams by well known mechanisms such as Target Normal Sheath Acceleration (TNSA) {\cite{Snavely2000, Wilks2001}}, Radiation Pressure Acceleration (RPA) {\cite{Esirkepov2004}} and others \cite{Borghesi2019} beneficial to isotope production \cite{Nemoto2001}, positron emission tomography \cite{Santala2001}, ion beam microscopy \cite{Merrill2009}, Particle-Induced X-ray Emission (PIXE) \cite{Mirani2021} as well as inertial confinement fusion \cite{Roth2001}. The mechanisms rely on the build up of large accelerating potentials which are also {the} source of ultra-strong electromagnetic pulses \cite{Consoli2020}. In particular, targets attain a strong positive net-charge due to laser-accelerated electrons that are able to escape the rising potential barrier \cite{Poye2018}. As a result of this, \si{\kilo\ampere}-level discharge pulses and return currents can be produced and propagate over the target surface \cite{Ehret2023a}. {Interest in these} effects is twofold: (i) both are {sensitive to} the total amount of charge that leaves the target and {therefore can be used as a passive diagnostic of the laser-target interaction}; and (ii) both allow to deliver all-optically generated \si{\kilo\ampere}-level \si{\nano\second}-duration current pulses that can be understood as {olive}{a} novel secondary source.

The monitoring of target discharge is an important aspect of ultrahigh intensity laser-solid interaction at high-repetition-rate. This paper presents an inductive current monitor as metrology for high-voltage pulses driven at high-repetition-rate. The measurement of return currents with inductive current monitors has been demonstrated previously in the regime of \si{\nano\second}-driver lasers with intensities \SIrange{1e14}{1e16}{\watt\per\square\centi\metre} \cite{Cikhardt2014}. We apply this technique to the characterization of discharge pulses driven by {high power} Ti:Sa systems {at relativistic intensities}.

Pulses of \si{\kilo\ampere}-level at \si{\nano\second}-duration pose a risk for electronic systems in {the} vicinity of the interaction \cite{Bradford2018,Dubois2018}, but they also {have} an application in the context of proton beam focusing \cite{Kar2016,Bardon2020} {olive}{and transient magnetic field generation \cite{Ehret2023a}}. We demonstrate {here} the stable generation of discharge pulses, with a clear perspective to obtain a novel high-repetition-rate source of \si{\kilo\ampere}-scale current pulses for future applications, {, e.g. on the field of electromagnetic compatibility (EMC) tests  \cite{A}, radio-location \cite{B}, military technologies \cite{C}, biology \cite{D}, and medicine \cite{E}.}

\section{Materials and Methods}

\begin{figure}
  \centering
  \includegraphics[trim={12cm 0.4cm 6.5cm 1.4cm},clip,width=0.67\columnwidth]{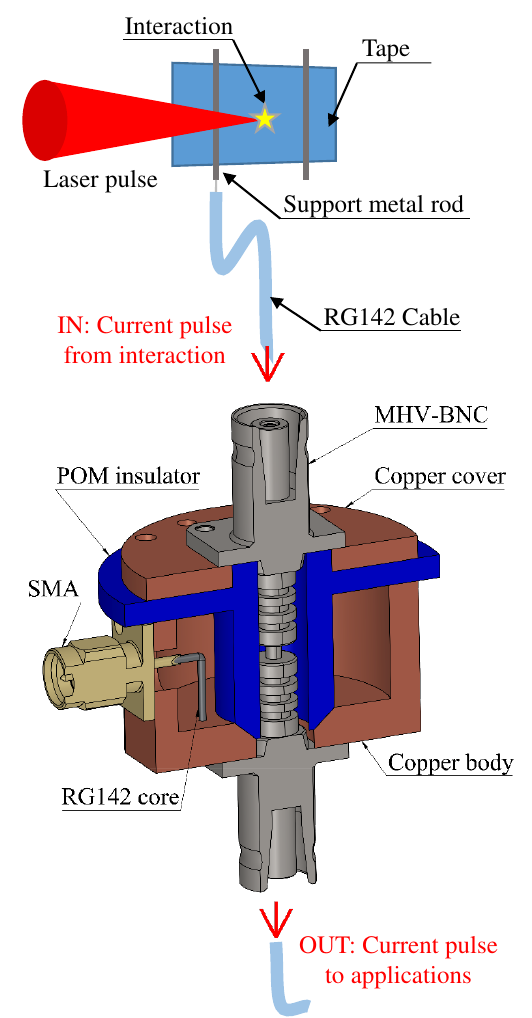}
  \caption{Cut of the Target Charging Monitor (TCM) with two opposite MHV-BNC connectors with soldered pins to pass through the {pulsed current issued by relativistic laser interaction in the top to the application side in the bottom. The TCM comprises} a solid copper body forming a cup; with a cylindrical top; both later of which are separated by dielectric material POM. The through current induces a magnetic field enclosed in the cylinder which causes an induced current to flow in a small squared loop formed by the core of a RG142 coaxial cable connected to an output SMA connector. {The current pulse itself is issued by the discharge of a solid tape target (top) and coupled into {olive}{one of} the support rods of the tape which are connected to a RG142 coaxial cable leading to the TCM. {olive}{The other support rod is isolated from the ground.}} }\label{fig:TCM}
\end{figure}

{The primary diagnostic used in this study was} a Target Charging Monitor (TCM) constructed based on the principles of an inductive current monitor \cite{Cikhardt2014}. The TCM measures the derivative of the current that streams through the device {as shown} in Fig.~\ref{fig:TCM}. {Key advantage of this metrology technique is its destruction-free nature. Current pulses are excited by laser-plasma interaction on a solid density target, transported through the TCM and can be applied after their characterization.}

Current pulses that pass through the TCM {device induce a magnetic field inside the cylindrical copper body} which causes an induced current to flow in a coil-shaped rod connected to an coaxial output. The calibration factor which relates the {time-integrated voltage} to the current is {\SI{-2.0 \pm 0.3 e9}{\ampere\per\volt}} (see Appx.~\ref{apx:TCMcalibration}). For this work, {current pulses are transported via RG142 coaxial cables and the circuit impedance is $Z=\SI{50}{\ohm}$. The through signal is terminated in the facility grounding. Note that cable {lengths} are measured with \SI{3}{\nano\second} FWHM voltage pulses: target and TCM are connected with a coaxial cable of \SI{9.6 \pm 0.2}{\nano\second} length; and TCM and grounding are connected with a coaxial cable of \SI{13.6 \pm 0.1}{\nano\second} length.} Induced signals are transported to a \SI{2}{\giga\hertz} oscilloscope and acquisitions are throughout corrected for the frequency dependent attenuation of circuit elements. {Circuit calibrations are done using a R\&S ZNH \SI{4}{\giga\hertz} vector network analyzer.} The effective bandwidth of the circuit is \SI{2}{\giga\hertz}.

Experiments for this work are conducted in the VEGA-3 laser facility at CLPU with {high-power Ti:Sa laser pulses} amplified to an energy $E_\mathrm{L}$ up to \SI{25}{\joule} per pulse {measured behind the compressor}. After compression {to a duration $\tau_\mathrm{L}$ of \SI{30}{\femto\second}}, the short laser pulse is transported in high-vacuum of \SI{1e-6}{\milli\bar} {via a $f=$ \SI{2.5}{\metre} off-axis parabola onto the target with a beam-transport efficiency of \SI{82}{\percent}.} The focal spot of $d_\mathrm{L}~=~\SI{12.8 \pm 1.9}{\micro\metre}$ full-width at half-maximum (FWHM) is maintained at constant size. {The energy on target is extrapolated from calibrations recorded at low-energy {and} the focal spot at high energy is estimated to be the same as for low-energy measurements. Note the large Rayleigh length of \SI{160 \pm 50}{\micro\metre}, for this work with a non-difraction limited focal spot}. For this experiment, \SI{33.9 \pm 1.5}{\percent} of the energy on target are within in the first Airy disk at low power. {In addition there are three hot spots aside the focus in the first Airy ring, containing in total as much as \SI{20 \pm 2}{\percent} of the energy on target with an average intensity of \SI{24 \pm 2}{\percent} of the main intensity.} The pulse duration is measured on-shot with a second-harmonic autocorrelator system that diagnoses the faint reflection from a thin {pellicle} positioned between parabola and focus.

The tape target system TaTaS-PW \cite{Ehret2023} {transported} aluminium tape of \SI{10 \pm 1}{\micro\metre} thickness, Kapton tape of \SI{89 \pm 9}{\micro\metre} thickness, tape of \SI{10 \pm 1}{\micro\metre} aluminium enforced with Kapton (Al-e-K) \cite{Ehret2023} and copper tape of \SI{7 \pm 1}{\micro\metre} thickness across the laser focal plane. {olive}{Tapes are \SI{12.5}{\milli\metre} wide stripes. The conductive \SI{5}{\milli\metre} diameter support metal rods which guide the tape are \SI{16}{\milli\metre} separated.} Solid-density targets are placed in the laser focus position and tilted by \SI{12.5}{\degree} with respect to the laser axis to avoid reflection back towards laser amplifiers. As the VEGA laser pulse shows no pre-pulses capable of inducing a transparency or breakdown of the target \cite{Volpe2019}, the main acceleration mechanism of charged particles is TNSA with the deployed laser and target parameters. In TNSA, a population of laser-heated {olive}{forward-directed} relativistic electrons escapes the target {olive}{after crossing its thickness} and the successive potential dynamics and electron-refluxing leads to the formation of sheath fields which are capable of accelerating ionized surface contaminants up to several tens of \si{\mega\electronvolt\per\atomicmassunit} \cite{Roth2016}. The TCM measures the return current towards the target and allows {us} to deduce the total target {charging} by relativistic electrons. {Note that the tape{olive}{'s only connection to the ground is one of the support metal rods. The other metal rod is floating with a total conductive length of \SI{95}{\milli\metre}. This} forces the return current to flow thought the {olive}{non-floating} support metal rod and successively in the core of the coaxial transmission line that incorporates the TCM. The shield of the transmission line is grounded with the chassis of the tape target system.}

{Numerical simulations are performed to compare experimental results to theoretical predictions. The laser-driven target discharge is simulated with ChoCoLaT-2 \cite{Poye2018} (see Appx.~\ref{apx:CoCoLaT}) - and the laser-absorption efficiency into hot electrons is studied with the particle-in-cell (PIC) code SMILEI \cite{SMILEI} (see Appx.~\ref{apx:PIC}).}

\section{Results and Discussion}
\begin{figure}
  \centering
  \includegraphics[width=\columnwidth]{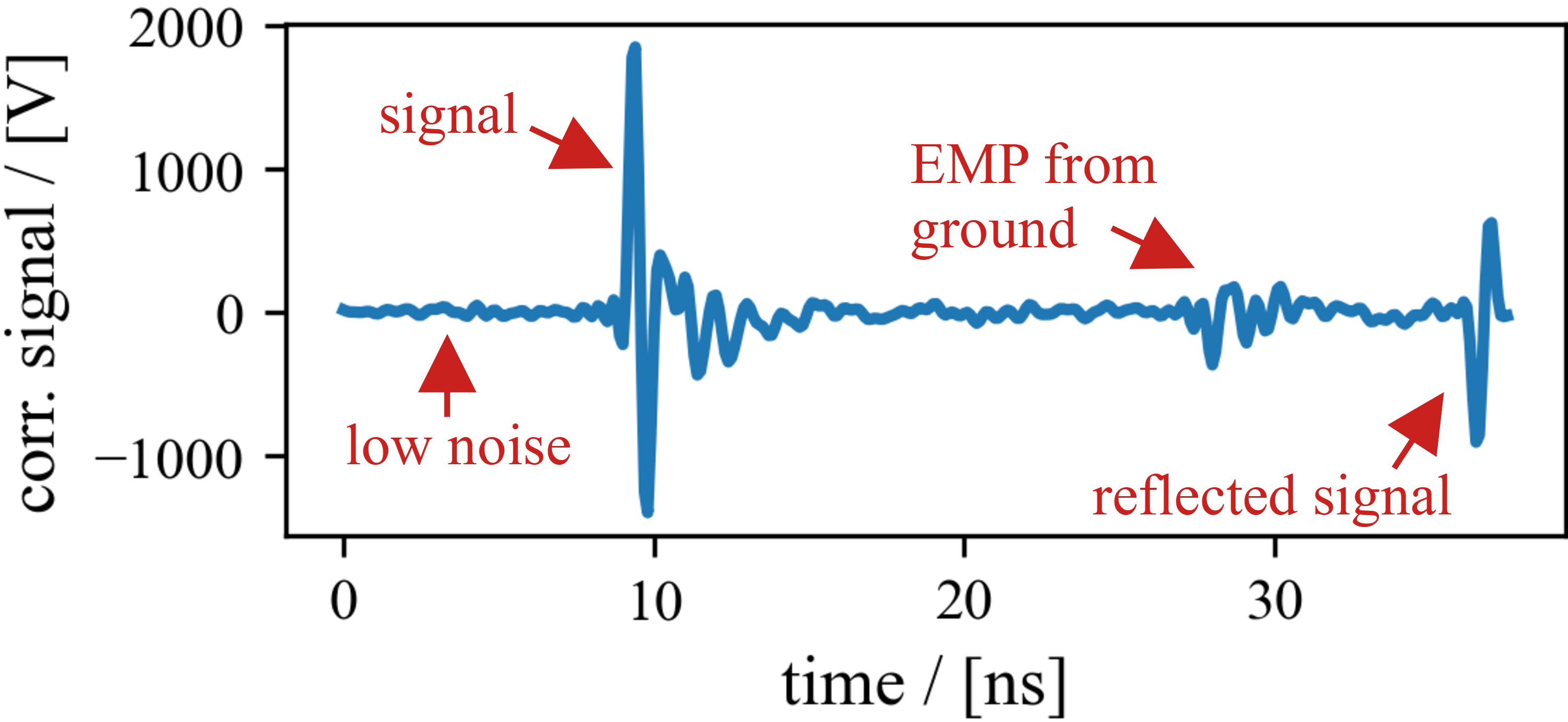}
  \caption{The circuit-corrected signal of the TCM for an aluminium target exhibits a clear positive peak for the rising edge of the positive current pulse. {It is preceded by a low-noise pedestal; and followed by pulses streaming from the grounding to the target: first EMP induced noise and second the reflection of the current pulse at the impedance miss-matched ground. Time-base {olive}{at TCM} relative to laser-arrival at \SI{\approx 0}{\nano\second}.}}\label{fig:dtIp}
\end{figure}

We first show results from {a single, representative shot on an aluminium target} to emphasize different aspects of the platform, and secondly study {the effect of changes of} laser and target parameters based on single-shot data and high-repetition-rate recordings.

The inductive TCM device measures the derivative $\mathrm{d}_\mathrm{t} I_\mathrm{p}$ of the pulsed current $I_\mathrm{p}$ streaming away from the target, see Fig.~\ref{fig:dtIp}. {Here, the laser pulse at \SI{22.5}{\joule} beam energy (after compressor) and a duration of \SI{30.4 \pm 0.7}{\femto\second} is} fired onto a \SI{10 \pm 1}{\micro\metre} thick aluminium target. {The laser impacts on the target at zero time seen from the TCM. The spatial distance between TCM and target is \SI{30 \pm 1}{\centi\metre} to ensure that the spherically expanding vacuum bound Electromagnetic Pulse (EMP) arrives at the device first. The EMP has no significant influence on the measurement as one does not notice noise in the pedestal leading to the signal.} The signal exhibits a first positive peak that detects the rising edge of the current pulse streaming through the device. This indicates a positive current pulse propagating from the target to the ground. {olive}{We measure the net negative charge escaping from the target -- the time between electron and ion escape is too short to be resolved.} {The measurement shows also a reflection of the current pulse that streams back from the imperfectly impedance matched grounding towards the target; and EMP induced noise that couples into the transmission line when the spherically expanding EMP reaches the grounding.}

\begin{figure}
  \centering
  \includegraphics[width=\columnwidth]{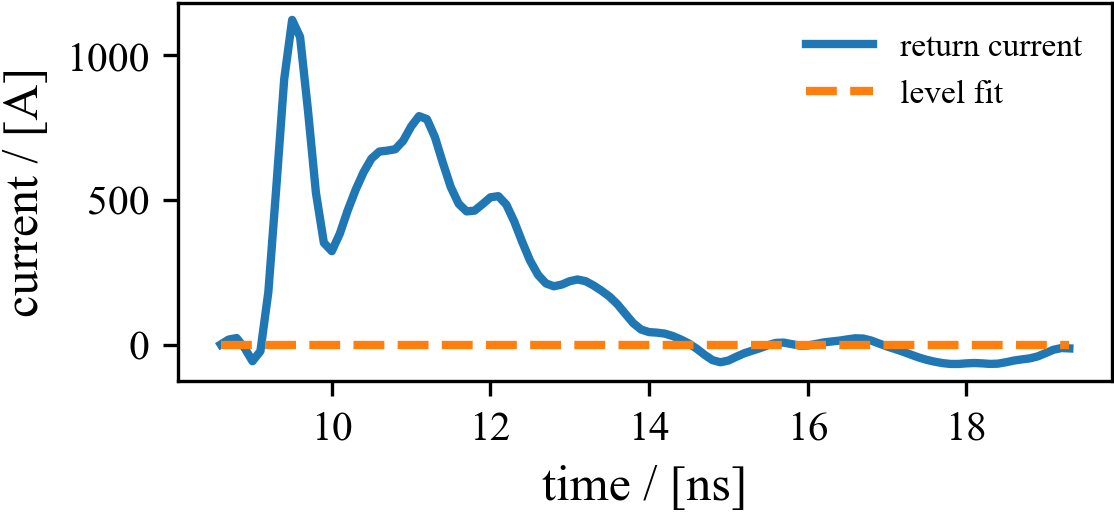}
  \caption{Current pulse (blue line) {from an aluminium target} retrieved by numerical integration from the derivative measured with the TCM. {A first short primary peak is followed by a superposition of peaks in a broad secondary peak.} {Time-base relative to laser-arrival at \SI{\approx 0}{\nano\second}.} The zero-level is controlled by comparison to a fit from before to after the current pulse (orange dashed line) -- here in good agreement.}\label{fig:Ip}
\end{figure}
After application of the instrument calibration, the {temporally-integrated} signal is shown in Fig.~\ref{fig:Ip}. The peak amplitude reaches \SI{1123 \pm 172}{\ampere}. The FWHM $\tau_\mathrm{d}$ of the narrow first peak is \SI{400}{\pico\second}. A broad second peak follows and decays slowly towards zero, which is reached after \SI{6}{\nano\second}. The first peak corresponds to the direct coupling of the discharge pulse into the {transmission line to the TCM}. {olive}{The difference between the shortest (direct to ground) and longest way (to opposite end of rod) from the interaction region to the exit of the grounding rod is equivalent to \SI{100}{\pico\second} at the speed of light. Capacitive effects may broaden the peak further.} The second peak most likely comprises multiple reflections across the {olive}{conductive tape} target.

\begin{figure}
  \centering
  \includegraphics[width=\columnwidth]{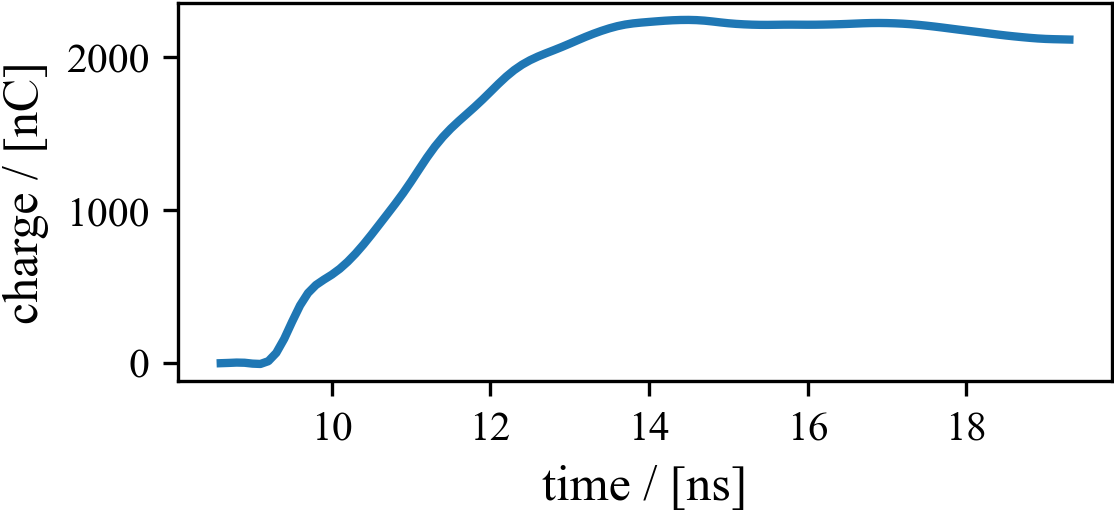}
  \caption{The transported charge from an aluminium target as obtained by numerical double-integration of the derivative measured by the TCM. The integral attains a plateau only slowly due to a slightly negative tail of the return current. {Time-base relative to laser-arrival at \SI{\approx 0}{\nano\second}.}}\label{fig:Qp}
\end{figure}
{Further temporal integration of $Z I_\mathrm{p}^2$ yields the transported energy $E_\mathrm{p}$ and $\int I_\mathrm{p} \mathrm{d}t$ yields the transported charge $Q_\mathrm{p}$.} {The total transported energy is \SI{67 \pm 7}{\milli\joule}.} {The energy conversion efficiencies from laser energy on target to current pulse energy results to $\chi_T = \SI{0.4}{\percent}$; and the energy conversion efficiency only accounting for laser energy encircled in the laser focus and relativistic-intensity hot spots calculates to $\chi_S = \SI{0.6}{\percent}$.} {The broad second peak contains a non-negligible fraction of the pulse energy in this configuration with \SI{43 \pm 7}{\milli\joule}.} The {temporally-resolved} transported charge $Q_\mathrm{p}$ is shown in Fig.~\ref{fig:Qp}. The laser extracts \SI{2.24 \pm 0.34}{\micro\coulomb} from the target. The first peak of the current accounts for less than a third of the transported charge and the slow decay of the second peak allows the integral to reach a plateau only slowly.

For a control of the accuracy of the numerical integrations, the zero-level is compared to a fit of both plateaus before and after the current pulse, shown as orange dashed line in Fig.~\ref{fig:Ip}. Here, the zero-level is maintained.

\begin{figure}
  \centering
  \includegraphics[width=\columnwidth]{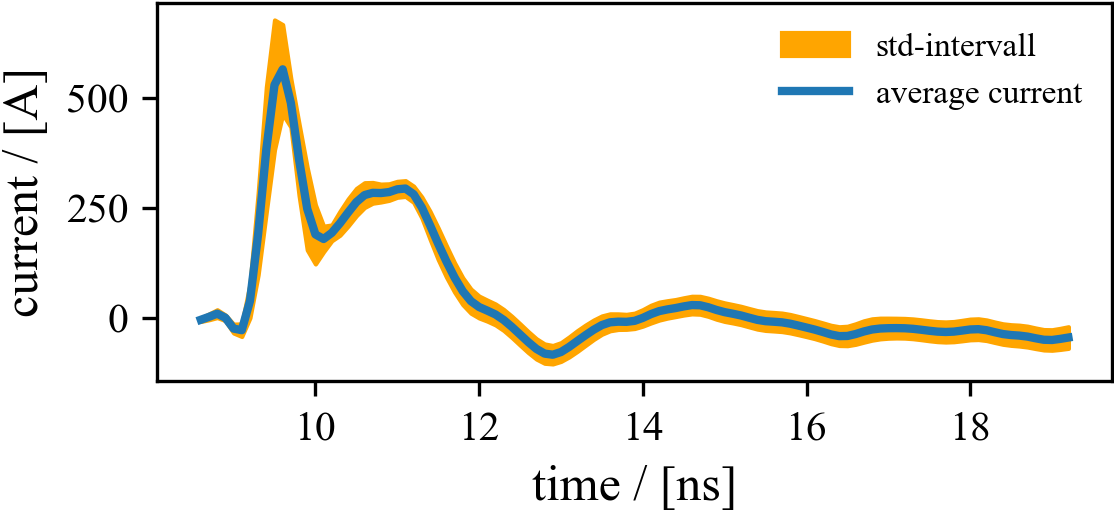}
  \caption{Average current and its standard deviation as obtained in $292$ laser shots of \SI{1.0 \pm 0.5 e20 }{\watt\per\square\centi\metre} at \SI{1}{\hertz} onto copper tape. {Time-base relative to laser-arrival at \SI{\approx 0}{\nano\second}.} {Multiple reflections across the conductive target yield a succession of multiple peaks.}}\label{fig:avg292Cu}
\end{figure}

Crucial for applications, the current pulse is reproducible {over hundreds} of shots {olive}{and} {is consistent with} {theoretical estimates}. A current of \SI{558 \pm 116}{\ampere} is obtained in $292$ shots at \SI{1}{\hertz} for laser shots of \SI{24.5 \pm 0.3}{\joule} at \SI{33 \pm 2}{\femto\second} onto copper tape of \SI{7 \pm 1}{\micro\metre} thickness. The average current {and its standard deviation} is shown in Fig.~\ref{fig:avg292Cu}. The {\SI{8}{\percent}} stability of the current {measurement} indicates a good shot-to-shot stability of laser and target parameters resulting in a reproducible discharge dynamics {and current production}. {The total transported charge} amounts to \SI{713 \pm 60}{\nano\coulomb} {and the current pulse energy is \SI{11 \pm 2}{\milli\joule}. The energy conversion efficiencies are lower compared to the ones for aluminium: $\chi_T^\mathrm{Cu} = \SI{0.05}{\percent}$ and $\chi_S^\mathrm{Cu} = \SI{0.09}{\percent}$}. 

{ChoCoLaT-2 simulations predict {\SI{720 \pm 75}{\nano\coulomb}} of target discharge} {when assuming \SI{68}{\percent} of the laser energy on target to be absorbed into electrons.} {Simulations take into account the experimental uncertainty for the pulse duration (\SI{33 \pm 2}{\femto\second}), \SI{6.8 \pm 0.3}{\joule} of laser energy within the first Airy disk and \SI{4.1 \pm 0.3}{\joule} distributed in three non-negligible hot spots with an average intensity of \SI{24}{\percent} of the main intensity.} {The absorption efficiency into electrons is consistent with PIC simulations, see Appx.~\ref{apx:PIC}.} {olive}{Such high values have been reported \cite{Ping2008}, depending on the presence of pre-plasma. If however the 2D PIC simulations should overestimate the absorption or no pre-plasma would be present, a typical \cite{Yu1999,Key1998} absorption of \SI{50}{\percent} would still lead to an agreement with overlapping uncertainty intervals.}

\begin{figure}
  \centering
  \includegraphics[width=\columnwidth]{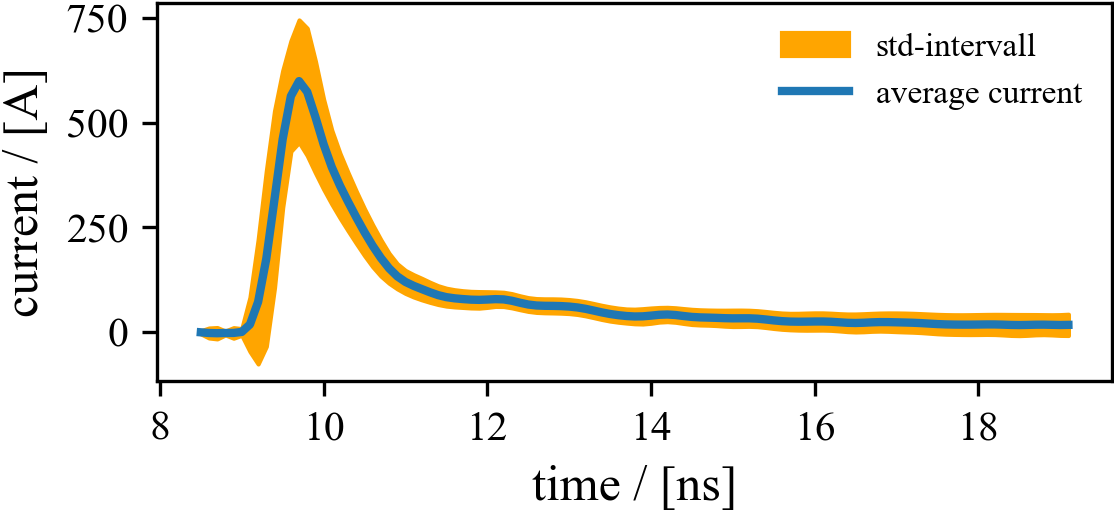}
  \caption{Average current and its standard deviation as obtained in $100$ laser shots of \SI{1.0 \pm 0.5 e20 }{\watt\per\square\centi\metre} at \SI{0.5}{\hertz} onto Kapton tape. {The dielectric target allows to produce single pulses. Time-base relative to laser-arrival at \SI{\approx 0}{\nano\second}.}}\label{fig:avg100K}
\end{figure}

A comparison of the metallic targets above to a dielectric target reveals the likely influence of target reflections and {shows how we can produce} single-peak current pulses. A current of \SI{597 \pm 153}{\ampere} is obtained in $100$ shots at \SI{0.5}{\hertz} for laser shots of \SI{22.9 \pm 0.2}{\joule} at \SI{33 \pm 1}{\femto\second} onto Kapton tape of \SI{89 \pm 9}{\micro\metre} thickness. The current {olive}{evolution (averaged over multiple shots)} is shown in Fig.~\ref{fig:avg100K}. The peak of \SI{960}{\pico\second} FWHM transports an average of \SI{934 \pm 190}{\nano\coulomb} {and has an energy of \SI{13 \pm 4}{\milli\joule}. The energy conversion efficiencies are comparable to the ones for copper: $\chi_T^\mathrm{K} = \SI{0.06}{\percent}$ and $\chi_S^\mathrm{K} = \SI{0.1}{\percent}$.} In comparison to metallic targets, the primary peak is broadened due to a reduced conductivity and secondary peaks from reflections are missing. {Multi-peak structures are indeed not expected to appear as reflections on tape ends do not occur, and reflections on other grounding stalks do not reach the signal transmission line with noticeable amplitude after the shot due to the low conductivity.} 
{Fewer} reflections on the target {may contribute towards the} reduced EMP emission that is generally observed for dielectric targets \cite{Consoli2020}.

\begin{figure}
  \centering
  \includegraphics[trim={0.4cm 0.4cm 0.2cm 0.3cm},clip,width=\columnwidth]{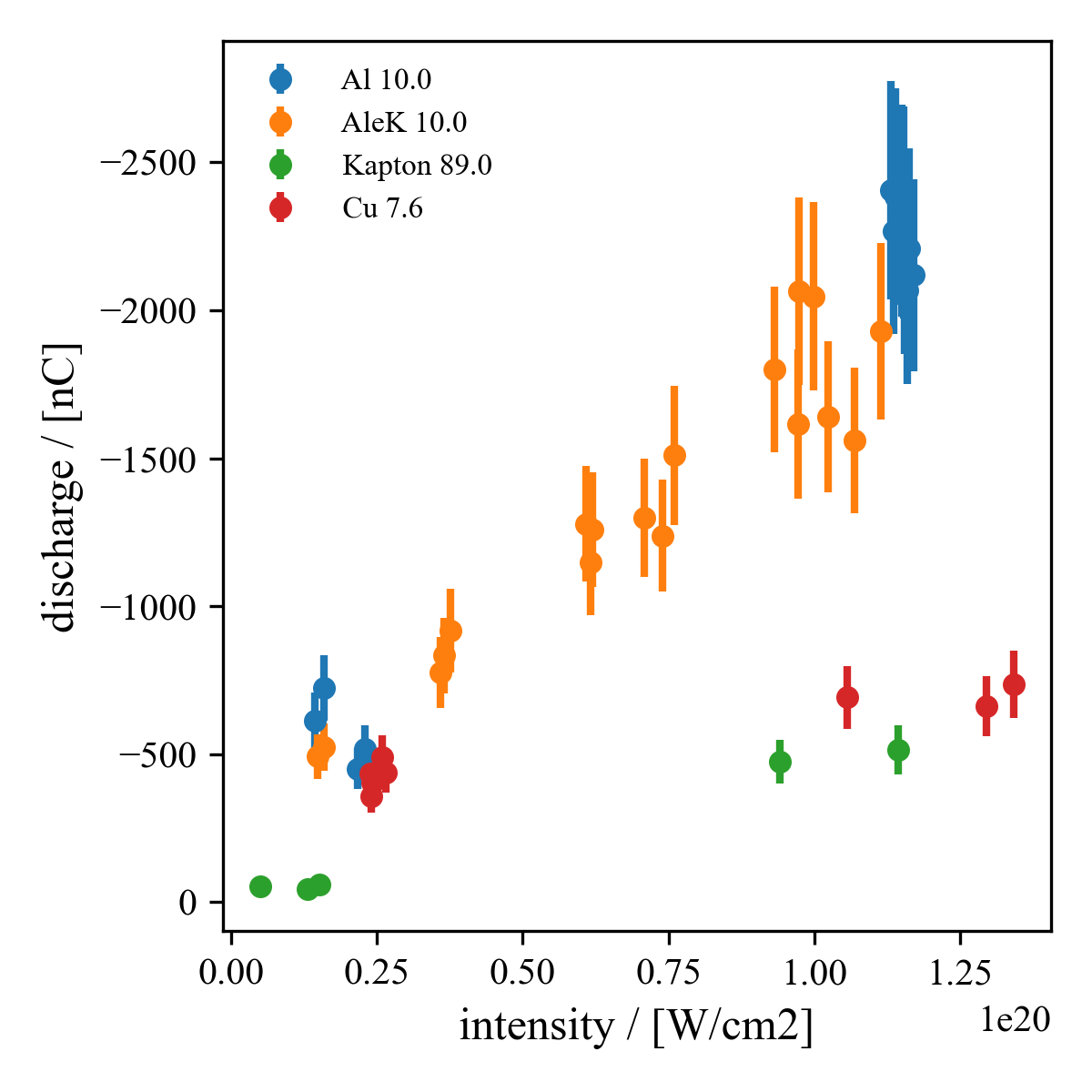}
  \caption{The total charge measured under variation of laser pulse duration and energy as well as the target material.}\label{fig:compare}
\end{figure}
It becomes clear that the current evolution and the amount of total charge varies considerably under variation of the target parameters. For a further parametric study on the variation of laser intensity, shots on aluminium tape, Kapton tape, tape of aluminium enforced with Kapton (Al-e-K) \cite{Ehret2023} and copper tape are compared in Fig.~\ref{fig:compare}. Most charge is ejected from aluminium targets, followed by copper and Kapton. Shots on aluminium reveal a linear relation between target discharge and intensity from \SIrange{2.0e19}{1.2e20}{\watt\per\square\centi\metre}. The platform allows for the production of tunable current pulses.

\begin{figure}
  \centering
  \includegraphics[width=\columnwidth]{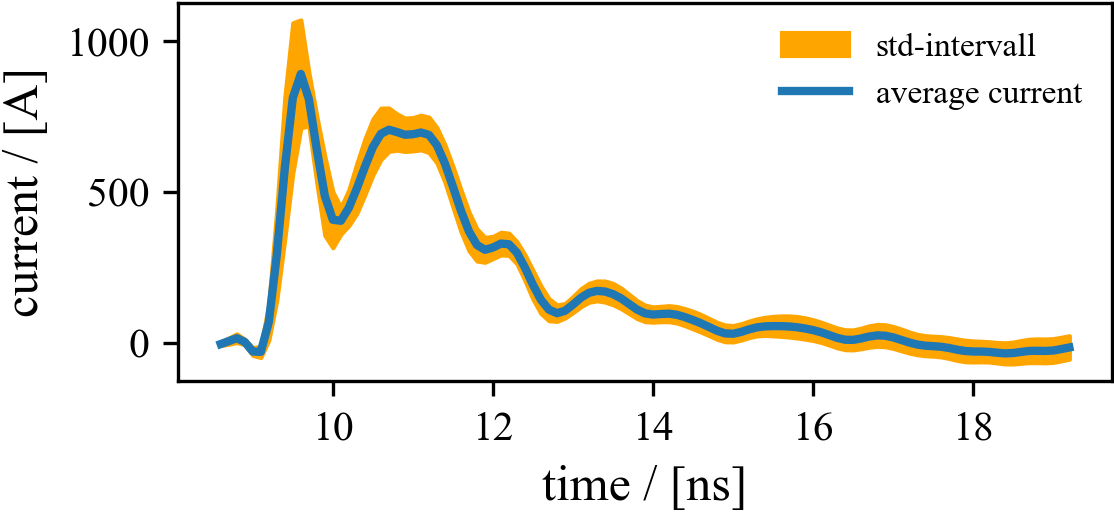}
  \caption{Average current and its standard deviation as obtained in $25$ laser shots of \SI{0.8 \pm 0.4 e20 }{\watt\per\square\centi\metre} at \SI{1}{\hertz} onto tape Al-e-K. {Time-base relative to laser-arrival at \SI{\approx 0}{\nano\second}.}}\label{fig:avg33AleK}
\end{figure}

Note {olive}{the geometry of tape Al-e-K: two \SI{5}{\milli\metre} wide strips of \SI{89}{\micro\metre} thick Kapton are glued on top of the aluminium tape at both its edges on the side facing the support metal rods. This} Kapton {reinforcement} of aluminium is at \si{\milli\metre}-distance from the interaction zone, {so} it does not change the total amount of ejected charge {(consistent with the experimental measurements in Fig.~\ref{fig:compare})}. The temporal shape of the current pulse can however be influenced by the Kapton enforcement for tape Al-e-K, see Fig.~\ref{fig:avg33AleK} (compared to Fig.~\ref{fig:Ip}). The first peak is lower, which is consistent with the {reduced} coupling to the grounding due to the presence of Kapton at the tape edges. As a result the secondary peaks in the tail are elevated for reasons of more charge in reflections.

{The characteristic parameters for all shot sequences are compared in Tab.~\ref{tab:AlAleK}. Current pulses from aluminium tape and reinforced aluminium tape are in a good agreement. Results for aluminium tapes exhibit higher current amplitudes when compared to copper targets due to a larger target discharge. }

\begin{table*}
\centering
\caption{{Comparison of current pulses from shots on aluminium tape (Al), Kapton reinforced aluminium tape (AleK), Kapton tape (Kapton) and copper tape (Cu). Laser energy measured after compressor, $N$ denotes the number of shots of the sequence, and $\chi_T$ is the ratio of energy confined in the current pulse to laser energy on target.}}\label{tab:AlAleK}
\begin{tabular}{c c c c | c c c | c c c c}
\hline
\multicolumn{4}{c}{laser pulse} & \multicolumn{3}{c}{target} & \multicolumn{4}{c}{current pulse}\\
\hline
energy & duration & rate & $N$  &  tape & material & thickness      & peak & charge & energy & $\chi_T$ \\
\hline
\SI{22. \pm 0.3}{\joule} & \SI{33 \pm 2}{\femto\second} & \SI{0.5}{\hertz} & $87$ & Al & Al & \SI{10 \pm 1}{\micro\metre} & \SI{982 \pm 185}{\ampere} & \SI{2.2 \pm 0.2}{\micro\coulomb}  & \SI{58 \pm 9}{\milli\joule} & \SI{0.32}{\percent} \\ 
\SI{21.9 \pm 0.3}{\joule} & \SI{37 \pm 4}{\femto\second} & \SI{1.}{\hertz} & 25 & Alek & Al & \SI{10 \pm 1}{\micro\metre} & \SI{809 \pm 210}{\ampere} & \SI{2.1 \pm 0.3}{\micro\coulomb}  & \SI{45 \pm 8}{\milli\joule} & \SI{0.25}{\percent} \\ 
\SI{22.9 \pm 0.2}{\joule} & \SI{33 \pm 2}{\femto\second} & \SI{0.5}{\hertz} & 100 & Kapton & Kapton & \SI{89 \pm 9}{\micro\metre} & \SI{597 \pm 153}{\ampere} & \SI{0.93 \pm 0.19}{\micro\coulomb}  & \SI{13 \pm 4}{\milli\joule} & \SI{0.06}{\percent} \\ 
\SI{24.5 \pm 0.3}{\joule} & \SI{33 \pm 2}{\femto\second} & \SI{1.}{\hertz} & 292 & Cu & Cu & \SI{7 \pm 1}{\micro\metre} & \SI{558 \pm 116}{\ampere} & \SI{0.71 \pm 0.06}{\micro\coulomb}  & \SI{11 \pm 2}{\milli\joule} & \SI{0.05}{\percent} \\ 
\hline
\end{tabular}
\end{table*}

\begin{figure}
  \centering
  \includegraphics[width=\columnwidth]{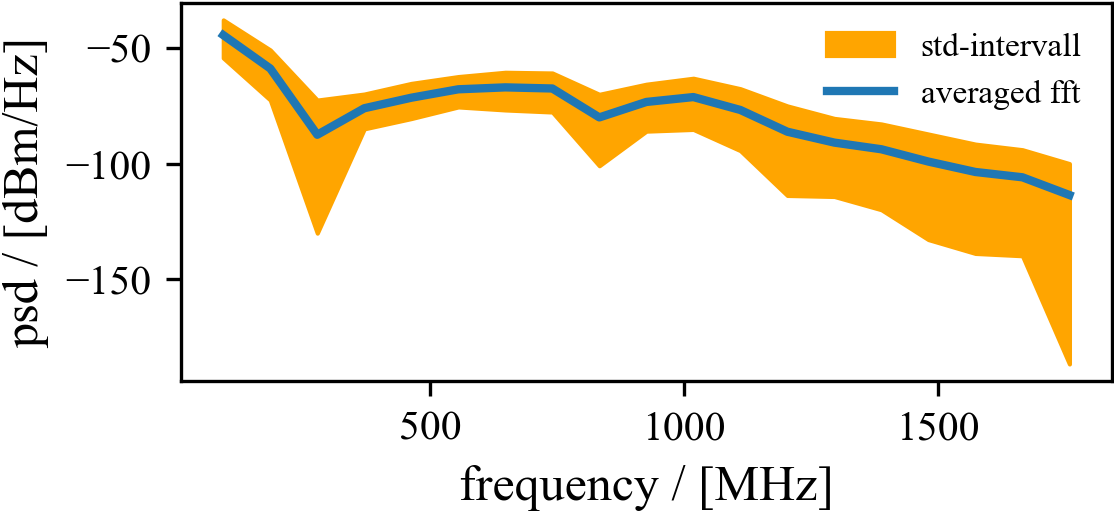}
  \caption{{Average power spectrum density and its standard deviation as obtained in $25$ laser shots of \SI{0.8 \pm 0.4 e20 }{\watt\per\square\centi\metre} at \SI{1}{\hertz} onto tape Al-e-K. Time-base relative to laser-arrival at \SI{\approx 0}{\nano\second}.}}\label{fig:fftIp}
\end{figure}

{The bandwidth of the current pulse is large and allows for applications that require broadband pulses \cite{B}, see  Fig.~\ref{fig:fftIp}. Such pulses can be applied to steering antenna arrays or impulse radiating antennas to emit high power levels, i.e. in ground and subsurface radars for finding, recognition and reconstruction of moving objects.}

\section{Conclusions}

We {report} the first generation and characterization of short pulsed \si{\kilo\ampere}-scale currents induced by high-power relativistic laser interaction at a high-repetition-rate. {The pulses with several \SI{100}{\pico\second} FWHM show a \SI{< 10}{\percent} stability {in} amplitude and a high energy conversion efficiency up to the order of \SI{1}{\percent} from laser energy to pulse energy.} {Optimization of the energy conversion efficiency will be possible by optimizing the target discharge based on existing theoretical models {olive}{\cite{Poye2018}}, as {the experimental data appears to agree well with simulations}. Current pulses} can be tailored by modifying the target: {experimental data shows that} the return current to metallic targets is broadened due to reflections across the target, whereas the use of dielectric targets removes those reflections leading to the generation of an overall shorter pulse peak.

{The highest charge of \SI{2.2 \pm 0.2}{\micro\coulomb} is produced with aluminium targets, followed by Kapton targets with \SI{0.93 \pm 0.19}{\micro\coulomb}, and copper targets with \SI{0.71 \pm 0.06}{\micro\coulomb}.}

A direct application of such pulses can be the inductive generation of pulsed strong magnetic fields in small volumes. The pulse fills a {olive}{solenoid} if $\tau_\mathrm{d} \times c = 2\pi \times r_\mathrm{c} \times N_\mathrm{c}$, with the speed of light $c$, the radius of the coil $r_\mathrm{c}$ and $N_\mathrm{c}$ revolutions. Then the induced magnetic field in the coil centre attains $B_\mathrm{c} = \mu_0 I_\mathrm{d} c \tau_\mathrm{d} / 2\pi r_\mathrm{c} l_\mathrm{c}$, with vacuum permeability $\mu_0$ and the length of the {olive}{solenoid} $l_\mathrm{c}$. The measured pulse of \SI{1.1}{\kilo\ampere} amplitude and \SI{400}{\pico\second} FWHM is apt for the generation of {olive}{\SI{11}{T} when using \SI{1}{\milli\metre} diameter coils of \SI{5}{\milli\metre} length, corresponding to $40$ revolutions}. Such magnetic fields can be used for the tailoring of \si{\mega\electronvolt\per\atomicmassunit} ions \cite{Kar2016,Bardon2020}, i.e. laser-accelerated ion beams. They are also relevant for magnetization of secondary samples \cite{Sakata2018,GPC2022,Walsh2022} {olive}{if further temporally stretched}, i.e. as seed fields in the context of magnetized implosions towards nuclear fusion.

Pulses of \SI{1.1}{\kilo\ampere} in the $50~\Omega$ circuit correspond to pulsed voltages of \SI{55}{\kilo\volt}, e.g. applicable to uni-polar nanosecond-pulse dielectric barrier discharge for producing non-thermal plasma at atmospheric pressure \cite{Tao2008}; or the research of effects of \si{\nano\second} and sub-\si{\nano\second} pulses on biological cells \cite{Sanders2009,Napotnik2016,Greenebaum2022,Porcher2023}.


\section*{Author Contributions}

The author contributions are as follows: ME, IMV performed the data acquisition, curation and analysis; ME wrote the first draft of the manuscript; JC, PB, ME, TB commissioned the device at PALS; JLH organized the beamtime at CLPU; DL, RHM managed implementation of the device;  ME, DL, JIA, RL contributed to conception and design of the study; all authors were involved with underlying experimental work; all authors contributed to manuscript improvement, read, and approved the submitted version.

\section*{Acknowledgements}

This work would not have been possible without the help of the laser- and the engineering teams at CLPU and PALS. Special thanks to the workshops of CLPU and PALS. This work received funding from the European Union’s Horizon 2020 research and innovation program through the European IMPULSE project under grant agreement No 871161 and from LASERLAB-EUROPE V under grant agreement No 871124; {and Euratom Research and Training Programme (Grant Agreements No. 633053 and No. 101052200 — EUROfusion)} {as well as from the Grant Agency of the Czech Republic (Grant No. GM23-05027M)}{; and Grant PDC2021-120933-I00 funded by MCIN/ AEI / 10.13039/501100011033 and by the “European Union NextGenerationEU/PRTR”.} The work was supported by funding from the Ministerio de Ciencia, Innovación y Universidades in Spain through ICTS Equipment grant No EQC2018-005230-P; further from grant PID2021-125389OA-I00 funded by MCIN / AEI / 10.13039/501100011033 / FEDER, UE and by “ERDF A way of making Europe” by the “European Union”; and in addition from grants of the Junta de Castilla y León with No CLP263P20 and No CLP087U16. {olive}{The views and opinions expressed herein do not necessarily reflect those of the European Commission.}

\section*{Data Availability Statement}

The raw data and numerical methods that support the findings of this study are available from the corresponding author upon reasonable request.

\appendix{}

\section{TCM Calibration}\label{apx:TCMcalibration}

\begin{figure}
  \centering
  \includegraphics[width=\columnwidth]{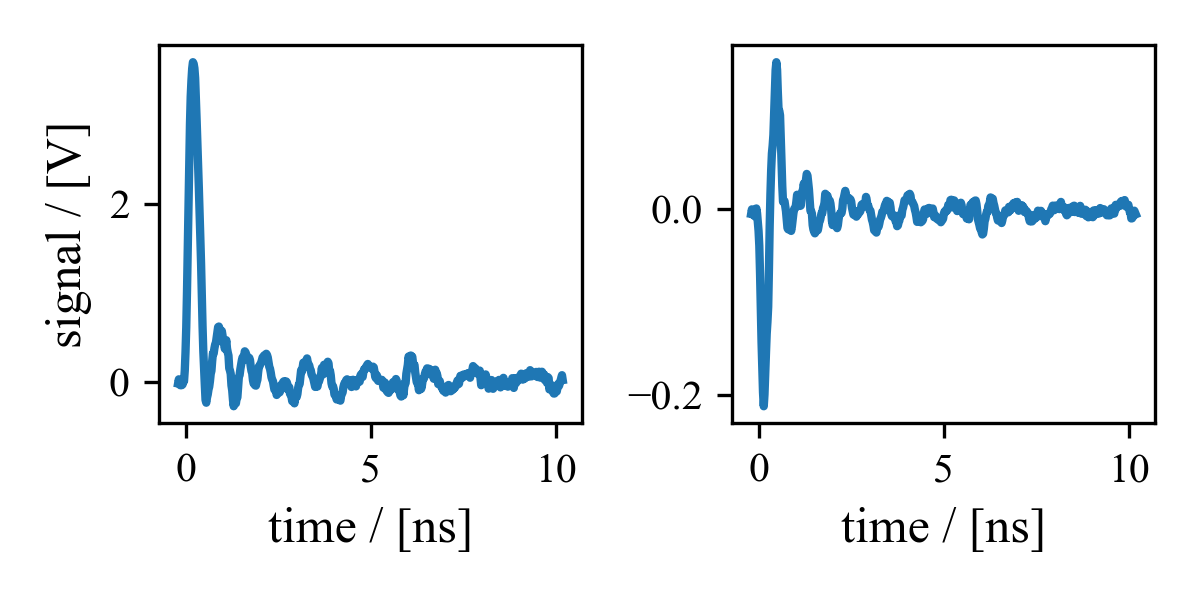}
  \caption{Through signal (left) and induced signal (right) corrected for attenuation of the respective circuit after the TCM.}\label{fig:thruandinduced}
\end{figure}

A pulsed high voltage supply {(\SI{500}{\pico\second} FWHM)} is used for the calibration of the Target Charging Monitor (TCM){. The voltage supply} is plugged to the top of the TCM. The through signal and the signal from the induced current are recorded on an oscilloscope of \SI{20}{\giga\hertz} bandwidth using calibrated coaxial cables in a \SI{50}{\ohm} circuit, see Fig.~\ref{fig:thruandinduced}. The current of the pulse can be naturally derived from the impedance of the circuit.

\begin{figure}[htb!]
  \centering
  \includegraphics[width=\columnwidth]{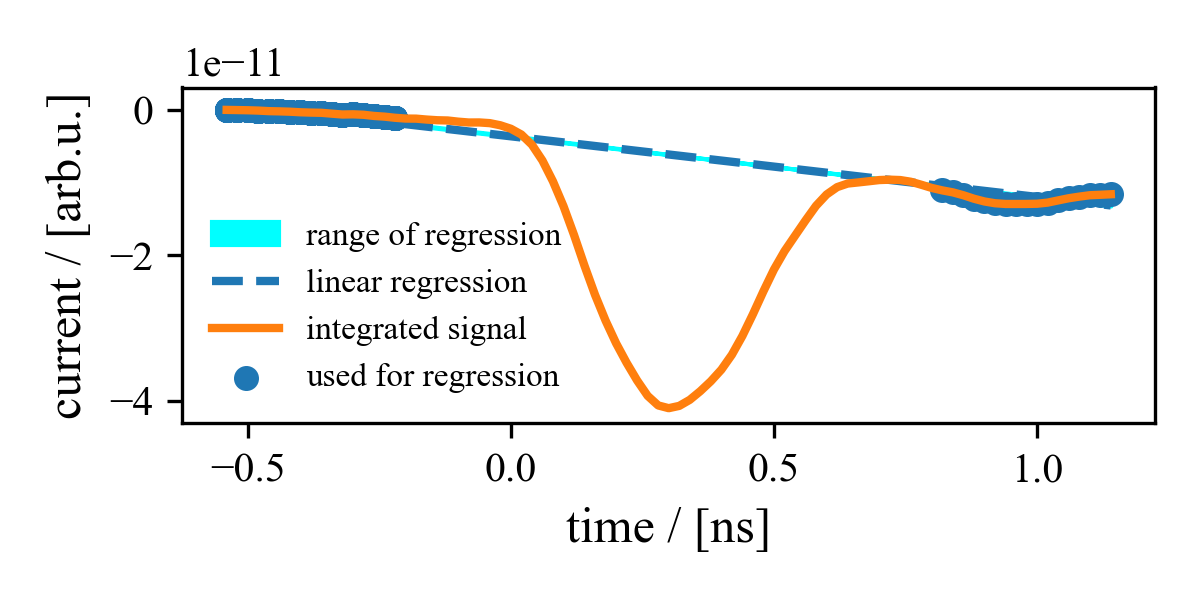}
  \caption{The integrated induced signal (orange line) shows a small offset after the pulse which might be due to numerical errors. Plateau regions before and after the peak are selected (blue dots) to fit a correction (dashed blue line) with respective uncertainty (cyan area). }\label{fig:integrateinduced}
\end{figure}

The induced signal is integrated numerically to derive the calibration factor between pulsed current and measured current, see Fig.~\ref{fig:integrateinduced}. The numerical integration may lead to what one observes as a change of the zero-level from before to after the peak. It is corrected for by fitting a zero-level with a linear regression. The difference between measurement and fit is fully taken into account in the following uncertainty estimates.

\begin{figure}
  \centering
  \includegraphics[width=\columnwidth]{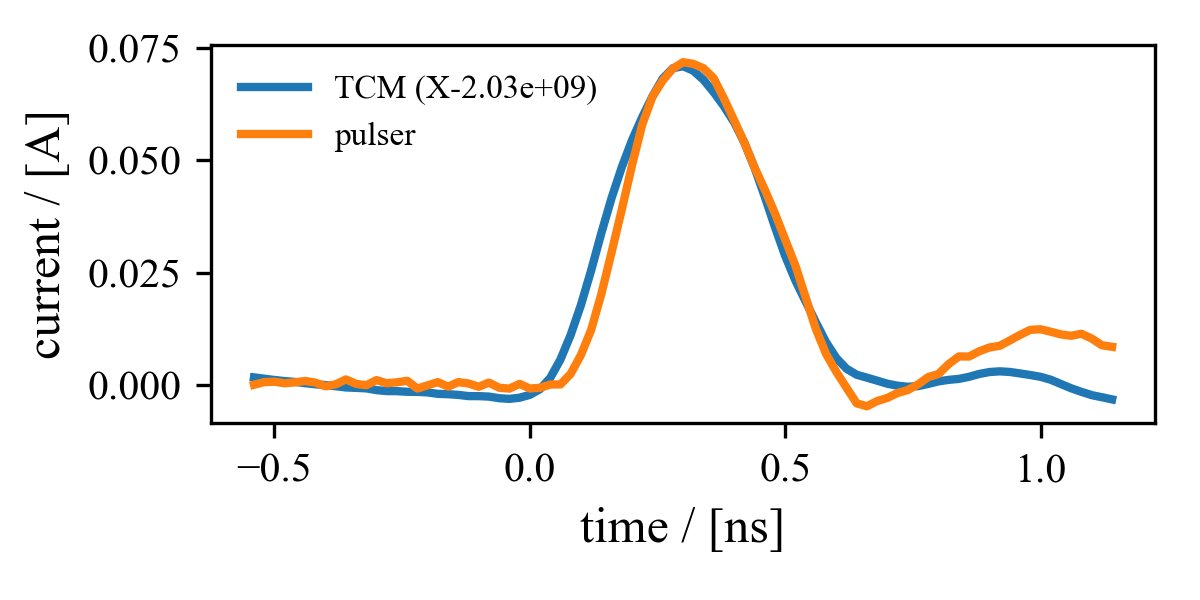}
  \caption{The integrated induced signal (blue line) is scaled to the pulsed through current (orange line) to obtain the calibration factor in units of \si{\ampere\per\volt}. }\label{fig:scaledintegratedinduced}
\end{figure}

The calibration factor which relates the integrated induced measurement in units of \si{volt} to the pulsed through current is obtained by fitting the base-corrected integrated induced signal to the through current, see Fig.~\ref{fig:scaledintegratedinduced}. One obtains a calibration factor of \SI{-2.0349 \pm 0.3117 e9}{\ampere\per\volt}.

\section{Discharge Simulations}\label{apx:CoCoLaT}

{The target discharge dynamics is studied using a detailed model of target charging in short laser pulse interactions \cite{Poye2018} that predicts the expected discharge current due to laser-heated relativistic electrons on a thin disk target. The successive electron escape is mitigated by the target potential, based on the driver laser parameters and the interaction geometry. The model takes into account the collisional cooling of electrons within cold solid density targets. The energy and time depending hot electron distribution function $f(E,t)$ describes electrons inside the target and evolves according to}
\begin{align}
\partial_t f (E,t) &= \frac{h_\mathrm{L}(E) \Theta(\tau_\mathrm{L} - t)}{\tau_\mathrm{L}}  -  \frac{f (E,t)}{\tau_\mathrm{ee}(E)} - g(E,t) \label{eq:distributionevolution}\\
h_\mathrm{L}(E) &\overset{!}{=} \frac{N_0}{T_0} \exp{\left[ - E/T_0 \right]} \\
N_0 &\overset{!}{=} \int f (E,0) \text{~d}E
\end{align}

{\noindent where $h_\mathrm{L}(E)$ is a constant exponential source of hot electrons, $\Theta (t)$ the Heaviside function limiting electron heating to the laser duration, $\tau_\mathrm{ee}(E)$ the energy dependent cooling time and $g(E,t)$ the rate of electron ejection from the target. The initial hot electron temperature $T_0$ depends on laser wavelength and pulse intensity \cite{Fabro1985,Beg1997,Wilks1992}; and $N_0$ is re-normalized to the energy balance $N_0 T_0 = \eta E_\mathrm{L}$ between the total energy of hot electrons in the target and the absorbed laser energy. Simulations require the conversion efficiency $\eta$ of laser energy to energy in the hot electron distribution, which is obtained by PIC simulations for this work (see Appx.~\ref{apx:PIC}).}

{The hot electron cooling time depends on target material properties such as mass density $\rho_\mathrm{t}$, mass number $A_\mathrm{t}$, atomic number $Z_\mathrm{t}$, and the hot electron
energy distribution that allows to calculate average speed $\left< v \right>_\mathrm{e}$ and energy $\left< E \right>_\mathrm{e}$. Its meticulous calculation is demonstrated in \cite{Poye2015} with an emphasis on cases relevant for this work.}

{With slight modifications to the source code, we enable the addition of hot-spot driven electrons to the electron distribution function. A section is added to construct an electron distribution based on the hot-spot energy and intensity, which then is added on top of the main electron distribution function.}

\section{PIC Simulations}\label{apx:PIC}

{The absorption of laser pulse energy into hot electron energy is studied for the case of shots on the copper target in a typical range of pre-plasma scale lengths. The absorption results to \SI{61}{\percent}} {for \SI{1}{\micro\metre} pre-plasma and \SI{78}{\percent}} {for \SI{3}{\micro\metre} pre-plasma. This range covers the absorption efficiencies required to reproduce experimental target charging by discharge simulations.}

{The 2D PIC simulation setup consists of a solid copper target irradiated by VEGA-3 laser system under an incidence angle of {\SI{12.5}{\degree}}. The laser is linearly polarized and has the following characteristics: a wavelength $\lambda$ of {\SI{800}{\nano\metre}}, a peak intensity of {\SI{7.5e19}{\watt\per\square\centi\metre}} (corresponding to a normalized field amplitude $a_0=5.9$), a pulse duration of {\SI{33}{\femto\second}} FWHM and a transverse waist of {\SI{13}{\micro\metre}}. The copper target is considered fully ionized, having a density $100\rm~n_c$, and a thickness of {\SI{7}{\micro\metre}}. The plasma density scale length is considered {\SI{1}{\micro\metre}} {and \SI{3}{\micro\metre} for the two distinct simulations}, having an exponential profile over a length of {\SI{20}{\micro\metre}} {, and being ablated from the initial target thickness}. The transverse width of the target is {\SI{40}{\micro\metre}}. At the rear side of the target we considered a thin layer of neutral protons of {\SI{70}{\nano\metre}} thickness and $10\rm~n_c$ density to simulate target contaminants. The simulation box has {\SI{80}{\micro\metre}} in the {longitudinal} direction and {\SI{40}{\micro\metre}} in the transverse direction. The cell length is $dx=dy={\SI{12.5}{\nano\metre}}$ and the number of particles per cell is 20 for each species. The particles are deleted while crossing the domain boundaries and the fields are absorbed. The simulations were performed with Simulating Matter Irradiated by Light at Extreme Intensities (SMILEI) \cite{SMILEI} on the cluster Supercomputación Castilla y León (SCAYLE) \cite{SCAYLE}.}

\end{document}